# Cancer systems biology in the genome sequencing era: Part 1, dissecting and modeling of tumor clones and their networks

Edwin Wang[1,2,*], Jinfeng Zou[1,3,4] Naif Zaman[1,5], Lenore K. Beitel[3,4], Mark Trifiro[3,4] and Miltiadis Paliouras[3,4]

1. National Research Council Canada, Montreal, Canada
2. Center for Bioinformatics, McGill University, Montreal, Canada
3. Lady Davis Institute, Montreal, Canada
4. Department of Medicine, McGill University, Montreal, Canada
5. Department of Anatomy and Cell Biology, McGill University, Montreal, Canada

* Corresponding to EW (edwin.wang@cnrc-nrc.gc.ca)
6100 Royal Mount Ave, Montreal, QC, Canada, H4P 2R2
Tel: 1-514-496-0914
Fax: 1-514-496-5143

**Abstract**

Recent tumor genome sequencing confirmed that one tumor often consists of multiple cell subpopulations (clones) which bear different, but related, genetic profiles such as mutation and copy number variation profiles. Thus far, one tumor has been viewed as a whole entity in cancer functional studies. With the advances of genome sequencing and computational analysis, we are able to quantify and computationally dissect clones from tumors, and then conduct clone-based analysis. Emerging technologies such as single-cell genome sequencing and RNA-Seq could profile tumor clones. Thus, we should reconsider how to conduct cancer systems biology studies in the genome sequencing era. We will outline new directions for conducting cancer systems biology by considering that genome sequencing technology can be used for dissecting, quantifying and genetically characterizing clones from tumors. Topics discussed in Part 1 of this review include computationally quantifying of tumor subpopulations; clone-based network modeling, cancer hallmark-based networks and their high-order rewiring principles and the principles of cell survival networks of fast-growing clones.

**Keywords:**
Genome sequencing, systems biology, molecular network, signaling network, cancer hallmark, tumor clonal network

# 1. Introduction

Advances in high-throughput technologies have made a great impact in revolutionizing medical research. Biology is becoming a data-intensive science through increasing use of "-omics" technologies. For example, whole-genome sequencing (WGS) has the potential to serve as a powerful and cost-effective diagnostic tool in the management of cancer. Tumor genome sequencing using WGS or WES (whole exome sequencing, i.e., sequencing of the coding regions of the genome) has provided huge amount of data for cataloguing genomic alterations of tumors including copy number variations (CNVs), insertions, deletions, and single-nucleotide variants (SNVs). It is expected that 10 000 tumor genomes will be completed using WES or WGS by the end of 2014. Today, sequencing a genome takes 1-2 days and costs ~$5 000. This means that genome analysis is now in the cost range of a sophisticated medical test such as magnetic resonance imaging. As sequencing technologies mature and costs are lowered, there has been an increase in the application of these technologies to tumor management.

Cancer is driven by changes in the genomes of cancer cells. Tumor genome sequencing efforts allows comprehensive cataloguing of genomic alterations. Thus far, these catalogues have shown that the key mutated genes and pathways that are altered in cancer are already well-known. EGFR, RAS, PI3K, P53, FGFR, MET and many other well-known cancer-driver mutating genes have been frequently rediscovered in many tumor genome sequencing studies [1-3]. Similarly, previously known key cancer-driving pathways such as RAS-pathway, PI3K-pathway, EGFR-pathway, MAPK-pathway and so on have been documented again and again in sequencing many types of tumors [1-3]. These results basically confirmed the major cancer biology knowledge generated via previous small-scale studies. The cataloguing activity does not provide many novel insights into the fundamental biology of cancer, and the catalogues alone may not unveil new cancer treatment strategies for cancer management or cure. However, the tumor sequencing activity does provide multiple types of data for tumors: mutation, CNVs, epigenetic and gene expression profiles.

Traditional cancer biology studies have focused on one gene or one pathway at a time. Given such complex data sets derived from tumor genome sequencing, it is possible to gain new insights by taking an integrative systems biology approach, instead of simply cataloguing genomic alterations. For example, a systems approach tends to develop new methods for integrating genomic alterations, functional information on genes and molecular networks to model cancer development, metastasis, and drug resistance at both a personal and systems level to identify new treatment strategies [4,5]. Although tumor genome sequencing provides few new insights into cancer biology, it allows dissection of subpopulations (clones) of tumors and generation of insights into tumor evolution. These results have profound impact on cancer systems biology and led researchers to reconsider how to conduct cancer systems biology studies. In this review, we will highlight our thinking about new directions for systems biology studies of cancer in the context of genome sequencing.

## 2. Quantifying tumor subpopulations via genome sequencing

In 1976, Novell [6] proposed that a single cell could randomly acquire a series of mutations that allow it to proliferate, then differently mutated cells, or clones could compete with each other, and one clone could outcompete others and finally form a tumor. Novell's hypothesis emphasized that only one clone (i.e., the one finally outcompetes others) was the most fit to survive and the other less-fit clones die out within a tumor. In late 80's, studies reported that a linear accumulation of specific genetic changes convert a normal epithelial cell into a tumor, in which a single clone is dominant [7-9]. These results strongly support Novell's hypothesis. However, later clinical treatments hint that a tumor is heterogeneous, i.e., a tumor could contain more than one clone. Recent tumor genome sequencing studies [10,11] confirmed that many distinct subpopulations of cells, or clones, co-exist in a tumor. Genome sequencing reveals the genetic record of their emergence over time and allows us to trace the divergence of a cell to form the different clones. By the time the cancer is diagnosed, one of these clones has become the dominant population in the tumor.

For quantifying subpopulation of tumors, new computational algorithms such as ASCAT [12] and ABSOLUTE [13] have been developed. These tools are able to: (1) infer the number of clones and their fractions within a tumor; (2) generate genetic scripts or profiles of genomic alterations including mutations and CNVs for each clone; and (3) infer the order of occurrence (timing) of the clones or the evolutionary tree of the clones. These tools opened a new window into the complexity of cancer, through reconstructing genetic networks of genomic alterations for clones of a patient's tumor, and then studying individual clones at a systems level. Some clones (early-occurring clones) occur in early stages of tumorigenesis, while some (late-occurring clones) appear in late stages. Based on selective growth advantages of the clones, we classify the clones into slow-growing and fast-growing clones. Slow-growing clones cannot form clinically detectable tumors on their own, whereas fast-growing clones are able to grow sufficiently to be clinically detectable as tumors without extra genomic alterations. In general, early-occurring clones are slow-growing clones, while late-occurring clones are fast-growing clones. Based on the studies of tumor genome sequencing, we proposed three genetic models of tumorigenesis using breast cancer as an example (Fig 1): (1) *the chromothripsis-driven model* [14,15]. Chromothripsis describes a process in which one or several chromosomes are shattered into hundreds of fragments in a single cellular catastrophe, and then the DNA repair machinery pastes them back together in a highly erroneous order. This process will generate gene amplifications/deletions on a massive scale. This is a fast track to form tumors which could contain very few clones. Only 2-5% of tumors may be generated in this fashion; (2) *the gradual mutation model*. This model suggests that cells accumulate mutations gradually and continuously to form slowing-growing clones first and then form fast-growing clones. The transition from slow-growing clones to fast-growing clones could be triggered by a genome duplication event (see Part 2 of this review [16]). One clone could have several direct daughter clones, thus, the clone populations form a family tree; and (3) the *stem cell model* [15]. This is similar to the gradual mutation model, but a stem cell clone could be formed at an early stage. A cancer stem cell clone generates new daughter cells, but with a limited and slow growth rate due to the unfavorable conditions in the microenvironment. These daughter cells further acquire new genomic alterations and then generate new clones. Clone populations in the tumor are also organized as a family tree, however, along the tree there might have a clone (stem cell clone) which could have significantly more branches than others.

Tumors formed via the gradual mutation or stem cell models could have one dominant fast-growing clone (e.g., representing 80-90% of a tumor volume), but most of the tumors could have a few fast-growing clones (the dominant one could represent 40-50% of a tumor volume). With these new understandings, and equipped with tools for dissecting and quantifying tumor clones, we should reconsider the strategies for conducting cancer systems biology studies, understanding the genetic/epigenetic underpinnings of human cancer, developing new insights into how to tackle this terrible disease, and finally new, more personalized treatments for cancer patients.

## 3. From tumor-based network modeling to clone-based network modeling

In the past, we studied a tumor as a whole entity. For example, we have generated enormous amounts of "omic" data for tumor samples in the past two decades. These data include gene microarray, RNA-seq, SNP, CNV and epigenetic profiles for all kinds of tumors. About ten years ago, a systems biology approach was applied to these data by focusing on modeling of molecular networks. However, the network approach has also considered a tumor as a single entity. In fact, all these data are a mixture of the profiles of multiple clones. Without dissecting clones, it is a black box for the number of clones and the frequency of each clone within a tumor. A tumor-based network thus describes 'crude' or inaccurate molecular interactions/regulatory relationships within the cancer cell. Therefore, tumor-based network modeling efforts have captured a large amount of noise, although some strong signals could be uncovered [17,18]. Without quantifying tumor clones, network modeling missed many important features of tumors, such as heterogeneity, which could be derived from clonal backup (i.e., within a tumor, when a clone is killed by a drug, another clone could overtake it and make the tumor recurred). However, at present, clinically, a drug only targets one clone within a tumor. Thus, tumor-based network modeling is ineffective in both uncovering fundamental insights into cancer biology and providing clues for better treatment of cancer patients.

Even knowing these drawbacks, in the past we have had to model cancer cells using tumor-based networks due to the lack of means of dissecting clones from tumors. Today, tumor genome sequencing provides an opportunity to model cancer cells by constructing and modeling networks for individual clones. By doing so, we could overcome the drawbacks of tumor-based network modeling and provide more accurate and comprehensive understanding of tumors, for example, we could model not only each clone's network, but also the potential functional interactions between clones within a tumor. Mostly importantly, clone-based network modeling could capture the features of individual clones (e.g., some clones have aggressive/invasive features, while other clones have no features of metastasis, but just grow rapidly). In this regard, clinical outcomes and drug interventions could be much more accurately examined by modeling of the clones within a tumor. To gain insights into tumorigenesis, we could answer key questions such as what, why and how specific combinatory patterns of mutations/CNVs occur in fast- vs slow-growing clones and drive network functions for selective growth advantages. Although having advantages, clone-based network modeling has many challenges in terms of: (1) except for the mutations and CNVs profiles, it is still not possible to dissect other data types such gene expression profiles for each clone within a tumor; and (2) the current methods for dissection of clones within a tumor still need to become more accurate. The new technologies such as single-cell genome sequencing and RNA-Seq could address these questions by profiling of the tumor clones.

**4. From global network modeling to key cancer hallmark-based network modeling**

Tumor genome sequencing confirmed that cancer is a mutation disease. To become a cancer cell, a normal cell normally experiences multiple rounds of mutations and generates thousands of genetic mutations. It has been suggested that promoting genome instability is a key factor in tumorigenesis. In addition, one of the fundamental characteristics of cancer cells is that they exhibit uncontrolled growth and proliferation. To have this characteristic, a fast-growing clone could lock its state to allow unlimited cell proliferation which involves promoting cell survival, cell cycle, cell proliferation and blocking apoptosis. Consistently, we have found that a few key cellular processes (i.e., key cancer hallmarks) related to cell proliferation, cell cycle, apoptosis, cytoskeleton and genomic instability (genes are responsible for chromatin or modification of histones) are significantly enriched in tumor genomic alterations through conducting Gene Ontology enrichment analysis of 'driver-mutating genes' (i.e., the mutations are able to promote or drive tumorigenesis, while other mutations are passengers that confer no selective growth advantage) from over 2 000 tumor genomes, which represent unbiased genome-wide sequencing of 24 000 protein-coding genes.

In the past, most cancer network analysis and modeling studies focused mainly on global networks such as protein-protein interaction networks or signaling networks [4,5]. The networks were modeled either in a general or a context-specific condition, however, they represented all biological processes [4,5]. We previously proposed that it is necessary to model cancer hallmark-specific networks (Fig 2A) to better understand key cellular processes which are involved in cancer development and progression [5]. Li et al. [19] applied this strategy to develop a computational algorithm for identifying cancer prognostic biomarkers and successfully improved the performance. They treated the genes that belong to one cancer hallmark as a module and searched the best gene combinatory effects within a module (i.e., a hallmark). The cancer hallmark approach provides several advantages: (1) cancer cells have to undergo key cancer hallmark cellular processes, therefore, genes which are belonging to cancer hallmarks are most likely playing roles in tumorigenesis or metastasis. To find biomarkers, focusing on hallmark genes will avoid gene noise which could erode the robustness of biomarkers; (2) genes in the same module (i.e., cancer hallmark) are likely to be regulated together, and therefore, biomarkers derived from these genes could have high robustness; and (3) core cancer hallmark processes are common processes for many tumor samples, therefore, they are likely to be the downstream components in genetic networks. In general, multiple upstream genetic pathways could trigger a limited number of key, common cancer hallmark processes. Thus, modeling of key cancer hallmark processes could lead to identifying relatively common cancer genes and key molecular mechanisms for different tumor samples. These key genes can be used as biomarkers or drug targets. Indeed, Li et al. showed that breast cancer prognostic biomarkers derived from the hallmark-based algorithm reached >90% predicting accuracy, whereas the biomarkers identified using other algorithms showed 59-79% predicting accuracies [19].

Cancer hallmark-specific networks could be further decomposed into network motifs, network modules and functional modules, which are the basic network 'building blocks' and bear important functional features [20,21]. These building blocks have been extensively studied and applied to cancer problems. For example, modules have been used as prognostic biomarkers [22,23]. Network motif patterns could predict that certain genes are biomarkers. Both miRNAs and oncogenes are preferentially targeted positive network motifs [24,25], therefore, miRNA decay or reducing the expression of Dicer (miRNA processing machinery) will predict cancer prognosis. Indeed, it has been shown that this is correct in many cancer types [26-29]. Adding other information to network/functional modules could improve biomarker discovery, for example, using the overlapping genes from single gene-based, gene-set (functional module)-based and network-based analyses on ovarian cancer data allows identification of more robust biomarkers [30]. Network modules have been also used for inferring function by applying a weighted co-expression network approach, which assigns a score to a pair of genes with respect to their co-expression frequencies across different cancer datasets or normal tissues [31]. In addition, network modules have been used to present biological pathways. Pathway enrichment analysis using weighted Kolmogorov-Smirnov statistics (PWEA), which considers weights of genes in a given pathway based on the shortest distances and correlation coefficients with other genes in the pathway, has been used for identifying key modules/pathways for diseases [32]. Mutual Exclusivity Modules (MEMo), which are based on somatic mutation, copy number and gene expression data of tumors, have been used to find onco-pathways [33].

A number of cancer hallmark-based network modeling and network module studies [34-37] have been conducted recently. Together with these examples, the key cancer hallmark processes highlighted by tumor genome sequencing emerge as the priority network modeling tasks. By modeling these hallmarks specifically in clone-based networks, we expect that principles for network regulation and rewiring, and key network modules that regulate specific hallmarks could be unveiled in the fast- and slow-growing clones. Because the current mutation cataloguing activity is insufficient to guide the development of more effective approaches for managing cancer patients, a better understanding of these hallmark-based networks is one of the most urgent tasks for cancer biology.

## 5. Uncovering high-order rewiring principles of cancer hallmark-based network interactions

Cancer hallmark-based networks could themselves be further organized into a high-order network structure to create core cellular processes that are co-regulated and functionally coordinated or enhance the cellular function of the networks. Different rewiring strategies could influence the outcome of the networks. There are many ways to connect two or more isolated networks. For example, a genomic alteration of a gene in one network (e.g., Gene A in a cell cycle network ) could gain a capability to regulate another gene or other genes in another network (e.g., Gene B of an apoptosis network is now regulated by Gene A of the cell cycle network , because of Gene A mutation). In this case, the two networks are connected. Another network connection possibility is that a gene, which does not belong to the two isolated networks, becomes mutated or amplified so that that gene could regulate genes in both networks. Currently, it is largely unknown how high-order rewiring principles (i.e., different linking strategies) affect the performance of the linked networks.

Recently, a game theory study of such a network problem showed that the properties of the nodes that are linked together often determines which network claims the competitive advantage [38]. Different ways of rewiring will generate different effects of two networks (cooperating or competing). A game theory-based method has been developed for analyzing how one network can maximize the cumulative benefit it gains when adding connective edges to other networks. Using this method, we could explore how different high-order rewiring strategies affect the benefit of gain or loss through interconnectivity. Translating these effects into a biological language, the method will help to judge how different high-order rewiring strategies affect the promotion, inhibition or coordination of the cellular functions of the two connected networks. There are multiple strategies for how edges can be added between two initially distinct networks (Fig 2B) and then generate beneficial gains for the networks. A high-ranked node in Network A could connect with a high-ranked node or a low-ranked node in Network B, while low-ranked nodes from both networks could also be linked. Connecting the low-ranked nodes typically enhanced the centrality (or importance) of the network that was initially 'stronger'. By contrast, linking high-ranked nodes generally increased the centrality of the 'weaker' network. This study implies that by choosing connective edges wisely, networks can maximize the benefit gained through interconnectivity. In this study, network centrality has been used as a measurement of benefit gain or loss. For biological networks, we are more interested in measuring cooperation, competition, enhancement, depletion, and complementation in terms of the cellular functions of networks. Furthermore, the study mentioned above only explored undirected links to connect two networks. We are more interested in directed links which represent activation or inhibition.

In summary, more interesting work could be explored for understanding of high-order rewiring principles for connecting different cancer hallmark-based networks. Cross-talk between modules/subnetworks is one example of the high-order network rewiring. A method of extracting a minimum connected network from a ranked list of genes for diseases from a network can reveal the crosstalk between subnetworks or pathways [39]. At the moment, this kind of study is still rare for biological networks.

## 6. Modeling of system constraints in fast-growing clones

Cancer cells need to acquire a selective growth advantage by random genomic alterations, so that they are able to form fast-growing clones, and ultimately, a tumor. Therefore, fast-growing clones will have one common feature – cell survival and proliferation – one of the fundamental cancer hallmarks. In historical cancer studies, more than 1 000 human cancer cell lines have been isolated from many kinds of tumors. In fact, these cancer cell lines represent fast-growing clones of tumors. Along with acquiring selective growth advantages, cancer cell lines also acquire fragile genes (for example, Gene A is not an essential gene in a normal cell, when that normal cell becomes a cancer cell, Gene A becomes an essential gene) due to their genomic alterations. Genome-wide RNAi screening of cancer cell lines suggests that cancer cell lines do not share common cancer-essential genes (i.e., knocking-down such genes leads to cancer cell die) [40,41]. It is clear that both cancer-driving genes (for example, mutating genes) and functionally essential genes are highly diversified, however, it seems that the cancer-essential gene sets are dependent on the cellular default settings of cell origins and the combinatory patterns of genomic alterations. Could we find rules about the relations between cancer drivers and cancer essential genes?

To answer this question, it is essential to model the genomic alterations and cancer-essential genes onto signaling networks. We previously proposed that signaling networks are important cellular communication platforms for driving tumorigenesis, based on the fact that a great fraction of historically discovered cancer mutation genes are signaling proteins [5,24]. Recent genome sequencing of more than 3 000 tumors provides an unbiased look at the emerging mutation data and also supports such a signaling-focused approach. Signaling proteins still are a large portion of the key recurrent mutation genes derived from tumor genome sequencing efforts. Further, a surveying of tumor CNVs showed that seven out of ten of the most frequently amplified genes among over 8 000 tumor genomes are kinases, including seven receptor tyrosine kinases (TKs), PDGFRB, EGFR, FGFR1, NTRK3 and ERBB2 [42]. These results suggest that signaling networks, especially TK-centered signaling networks are important for modeling of cancer development, progression and metastasis.

Constructing a human signaling network often relies on collecting signaling pathways identified in the past. Major signaling pathway databases include Pathway Interaction Database [43] and Reactome [44]. Recently, a computed high-resolution kinase-substrate network containing ~4 000 relations was constructed [45]. We have maintained the largest human signaling network containing >6,300 proteins and >62,000 signaling regulatory relations by manually curating and updating the network since 2006 [24,46]. In the current signaling networks, we included the data mentioned above and a recently curated tyrosine kinas signaling circuit dataset [47]. In addition, signaling network analysis and modeling has also been extensively conducted in previous studies [48,49].

To model the cancer cell survival network, Zaman et al. (submitted for publication) used 16 breast cancer cell lines which belong to two different molecular subtypes, luminal (non-aggressive) and basal (aggressive) subtypes (Fig 3). For each cell line, genome-wide RNAi, exome-sequencing and CNV data have been generated. Mapping these data onto the human signaling network, they constructed a cell survival-specific signaling network by highlighting the cancer-essential genes and potentially cancer-causal genes. They further identified subtype-specific survival signaling networks (~200 genes in each network) in which genes are recurrently used as essential genes or cancer drivers (i.e., genomic alterations) across their subtype's cell lines. Although the cancer-casual genes (or essential genes) are rarely shared between the cell lines within a subtype, they repeatedly hit ~200 network genes in a way that an essential gene in one cancer cell becomes a causal gene in another cancer cell and vice versa. In another words, the merged gene lists (merging of the essential genes and cancer-casual genes for a cell line) between cancer cell lines within a subtype are much more similar.

These networks highlight functional constraints and evolutionary convergence for the causal genomic alterations that drive cancer cell proliferation. Therefore, the networks suggest underlying mechanisms governing how genomic alterations cause cancer development and indicate the 'deterministic fate' for causal genomic alterations. Such a deterministic fate suggests that these networks have predictive power for subtype-specific behaviors. To demonstrate this point, it was shown that the mutation and amplification status of the highly connected genes in the networks significantly classified and distinguished 402 breast tumor samples. Such a classification allows to conducting a non-invasive approach (i.e., performing plasma genome sequencing [50]) to obtain the mutation and CNV profiles of the differentially network hubs, which, in turn could predict breast cancer, even the subtypes. To further demonstrate the predictive power of these networks, using a differential network biology approach [51], Zaman and colleagues (unpublished observations) showed that the networks can be used to predict subtype-specific drug targets. Most (~80%) of the predictions have been successfully validated in more than 50 breast cancer cell lines, thus, this analysis indicates that there are 'rules' for functional selection of causal genomic alterations which drive cancer survival for a given cancer subtype, and provides clues for cancer subtype-specific treatments. We hope that this approach could be used for multiple clones within a tumor so that targets for the clones within a tumor could be suggested. By doing so, a real personalized and effective treatment strategy could be taken for a patient.

**Acknowledgements**
This work is supported by Genome Canada and Canadian Institutes of Health Research. Some of concepts are beneficiated from the stimulating discussion in the AACR-NBTS-Cancer Systems Biology Think Tank.

**Figure Legends**

**Figure 1. Genetic models of tumorigenesis using breast cancer as an example.** (1A) *the chromothripsis-driven model*. Chromothripsis describes a process in which one or several chromosomes are shattered into hundreds of fragments in a single cellular catastrophe, and then the DNA repair machinery pastes them back together in a highly erroneous order. This process will generate gene amplifications/deletions on a massive scale; (1B) *the gradual mutation model*. This model suggests that cells accumulate mutations gradually and continuously to form slowing-growing clones first and then form fast-growing clones. The transition from slow-growing clones to fast-growing clones could be triggered by a genome duplication event. One clone could have several direct daughter clones, thus, the clone populations form a family tree; and (1C) the *stem cell model*. This is similar to the gradual mutation model, but a stem cell clone could be formed at an early stage. A cancer stem cell clone generates new daughter cells, which further acquire new genomic alterations and then generate new clones.

**Figure 2. Cancer hallmark-based networks.** (2A) Dissecting of cancer hallmark-based networks. For each cancer hallmark, its genes are often organized into a small network; (2B) Strategies for adding edges between two distinct networks. A high-ranked node one network could connect with a high-ranked node or a low-ranked node in the other network, while low-ranked nodes from both networks could also be linked. Connecting the low-ranked nodes typically enhanced the centrality (or importance) of the network that was initially 'stronger'. By contrast, linking high-ranked nodes generally increased the centrality of the 'weaker' network. Big and small circles represent high-ranked and low-ranked nodes, respectively.

**Figure 3. Modeling of systems constraints of cancer cell survival network.** Constructing cancer survival signaling networks of 16 breast cancer cell lines which belong to two different molecular subtypes, luminal (non-aggressive) and basal (aggressive) subtypes. For each cell line, genome-wide RNAi, exome-sequencing and CNV data have been integrated onto a human signaling network, then constructing subtype-specific survival signaling networks. Interestingly, in these networks, genes are recurrently used as essential genes or cancer drivers (i.e., genomic alterations) across their subtype's cell lines. These networks highlight functional constraints and evolutionary convergence for the causal genomic alterations that drive cancer cell proliferation. These networks correctly predicted cancer subtype-specific drug targets.

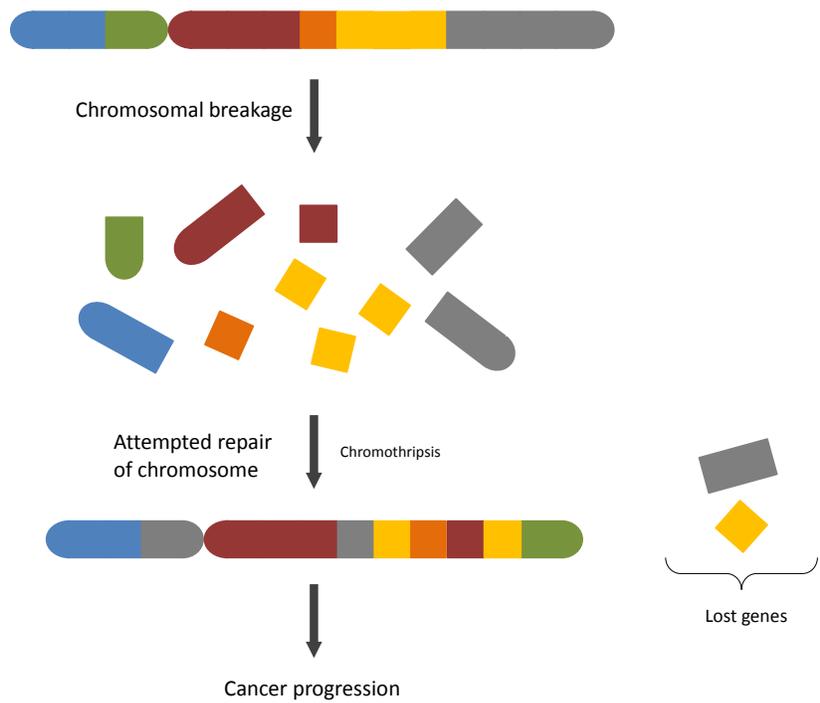

Fig 1A

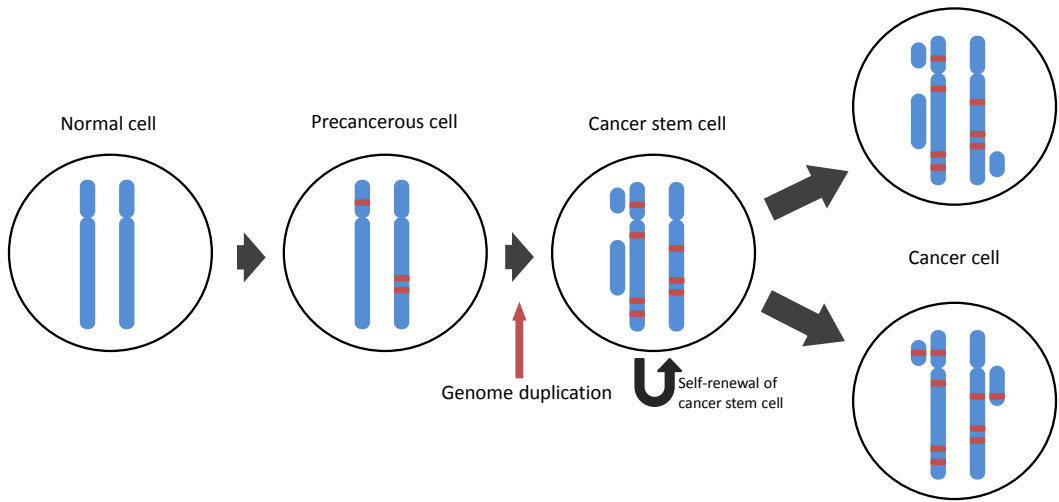

Fig 1B

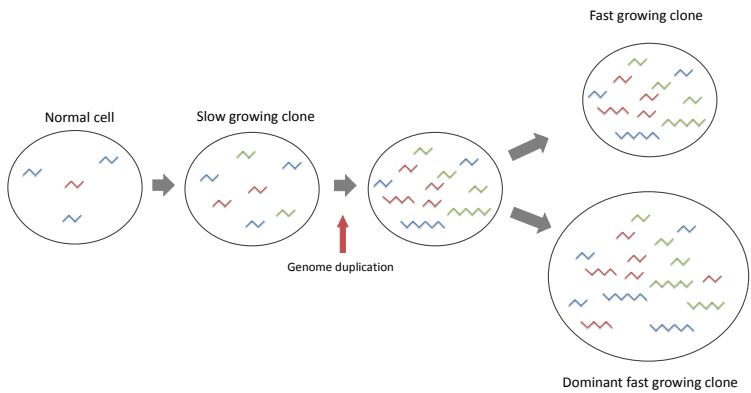

Fig 1C

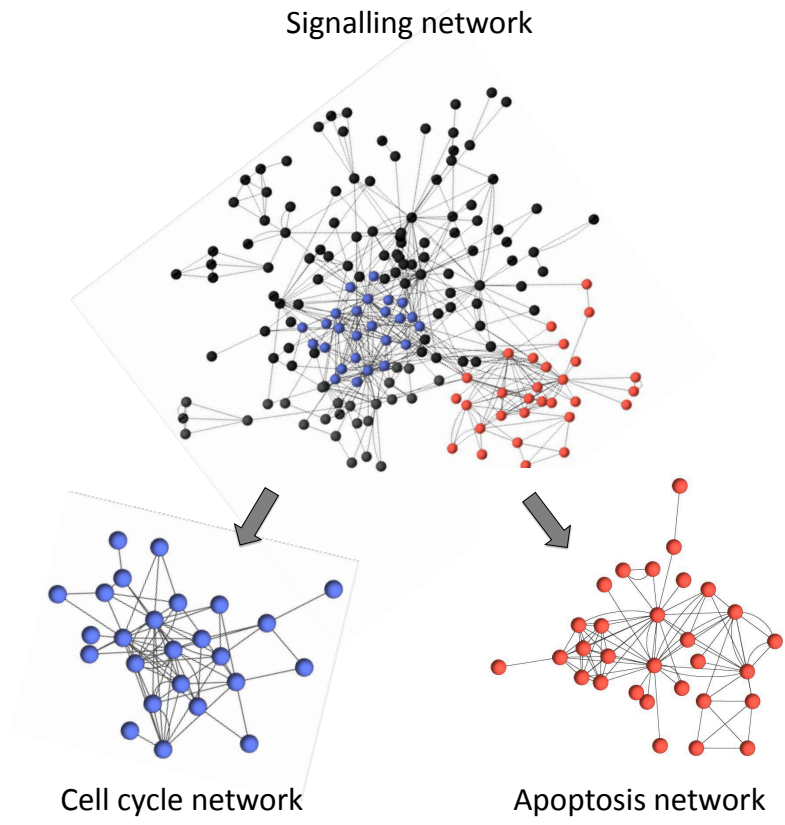

Fig 2A

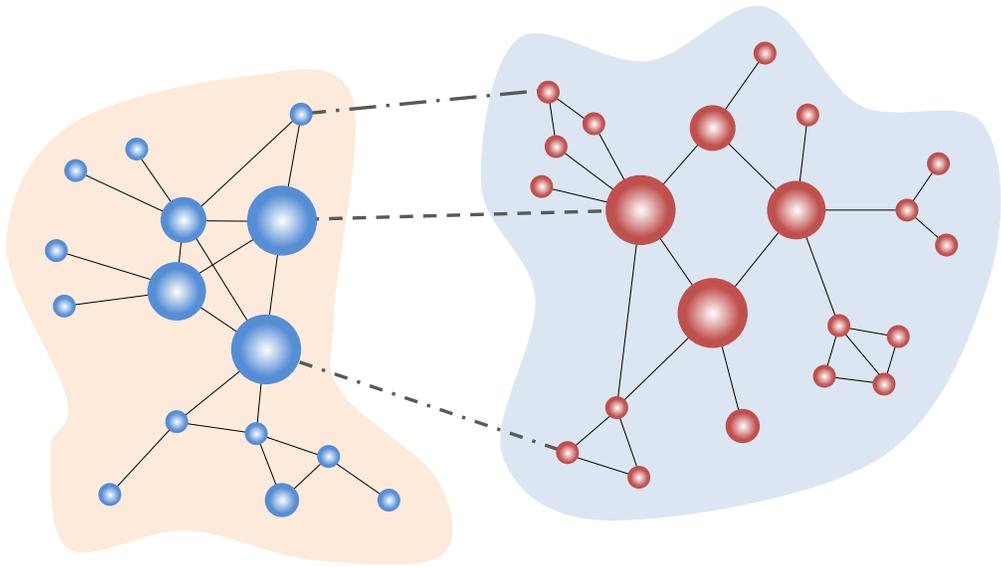

Fig 2B

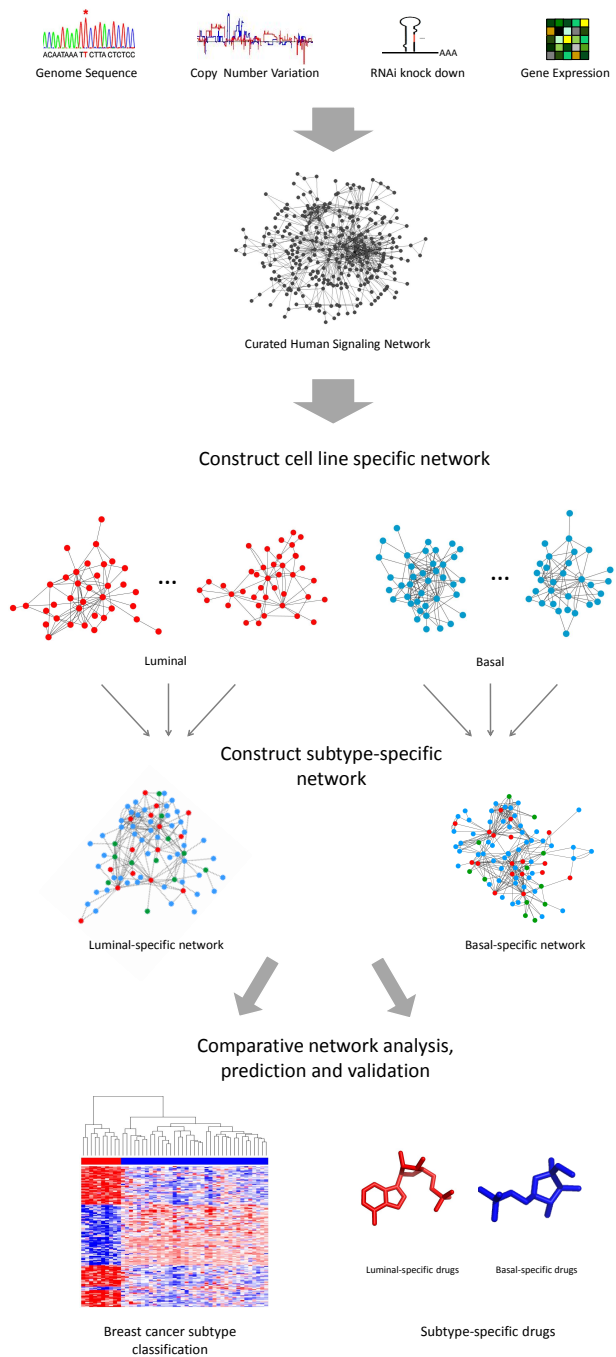

Fig 3